# Investigation of the Effects of Biodiesel Produced from Crambe Abyssinica Plant on Combustion, Engine Performance and Exhaust Emissions

Ayhan Uyaroğlu[1], Tolga Kocakulak[2*], Bilal Aydoğan[3]

[1]Selçuk University, Cihanbeyli High Vocational School, Konya, TURKEY

[2]Burdur Mehmet Akif Ersoy University, Technical Sciences of High Vocational School, Burdur, TURKEY

[3]Bandırma Onyedi Eylül University, Maritime of High Vocational School, Balıkesir, TURKEY

ayhan.uyaroglu@selcuk.edu.tr, tkocakulak@mehmetakif.edu.tr, baydogan@bandirma.edu.tr

**Abstract**

In this study, biodiesel fuel produced from crambe abyssinica plant using KOH and NaOH catalysts was mixed with standard diesel fuel and the effects on engine performance, combustion and emission were experimentally investigated. Standard diesel fuel and 6 different fuel mixtures were tested at 3.75, 7.5, 11.25 and 15 Nm engine loads. During the experiment, in-cylinder pressure data were specified depending on crank angle for each test fuel and engine load. In addition, measurements of HC, $NO_x$, CO and smoke emissions were carried out. With the obtained experimental data, parameters such as heat release rate, combustion stages, thermal efficiency, indicated mean effective pressure (imep), ignition delay (ID), ringing intensity (RI) and specific fuel consumptions (SFC) were calculated and evaluated in MATLAB/Simulink environment. The same concentrated crambe abyssinica with NaOH catalyst (CAN) blended fuel series has been found to have a lower BSFC value at all engine loads than the crambe abyssinica with KOH catalyst (CAK) blended fuel series. It was concluded that the highest thermal efficiency values were achieved with CAK B25 mixed fuel under all engine load conditions. It was concluded that the usage of standard diesel fuel is more prone to knock than other blends.

# 1 Introduction

Today, with the increase in the usage of motor vehicles, the risk of air pollution problems have reached significant levels [1,2]. In addition, the usage of internal combustion engines on motor vehicles cannot be dispensed with for many reasons. This situation has directed researchers to the production and use of more environmentally friendly renewable energy resources. At the basis of these renewable energy sources is biodiesel fuel that can be produced from vegetable oils and animal fats found in industry [3-8]. One of the most important advantages of biodiesel is that biodiesel does not require any extra equipment for its usage in diesel engines. Although biodiesel fuel on diesel engines causes some decrease on engine performance compared to standard diesel fuels, remarkable improvement on emission values can be observed [9-12]. Although biodiesel is produced from many edible oils, their social and economic sustainability is at risk and they cannot compete. In addition, it is predicted to negatively affect the hunger problem that occurs and may occur in underdeveloped and developing countries [13, 14]. Researchers have sought an alternative, inedible raw material to biodiesel production [15-17]. At this point, crambe abyssinica, one of these inedible raw materials, has been found to be suitable in many respects for biodiesel production [18].

Crambe abyssinica is an oil plant of the cruciferous family that naturally grown in Ethiopia, its height 1-2 m, it has white or yellow flowers and has higher heating value and oxidative stability advantages for biofuel production by comparison with the soybean biodiesel [19, 20]. Crambe abyssinica is an annual plant and is used in the machinery industry, oil industry and biodiesel production due to its high oil efficiency between 35% and 60%. Costa et al., [21] in his study, stated that they obtained oil between 26% and 34% from crambe abyssinica plant grown in different years in Portugal. The remaining part of the oil extracted crambe abyssinica plant is also widely used as a food product in livestock [22]. According to Zanetti et al. [23] study findings; growing cycle of crambe is 100 days in Northern Italy, 176 days in Southern Italy averagely. Wang et al. [24] have found the growing cycle of the crambe as 212–214 days and theoretical yield of 1485–5250 kg/ha in Chengdu area. Falasca et al. [20] have stated that the yield of crambe can vary 1125–1622 kg/ha in Russia and 450–2522 kg/ha in U.S.A. Zanetti et al. [23] have found that the mean oil content of dehulled seeds approximately 400 g/kg with independent of the location and varieties. On the other hand, they have investigated the three crambe varieties grown in two seasons in Italy in the sense that the major productive parameters and conversion efficiency. They stated the mean oil yield 986 (kg/ha) and 687 (kg/ha) in two seasons. Jasper et al. [25] have expressed that crambe has the the lowest cost of installation and conduct compared to sunflower, canola and soybean respectively. Fig. 1 is shown the development levels of crambe crop between the sowing and harvesting periods.

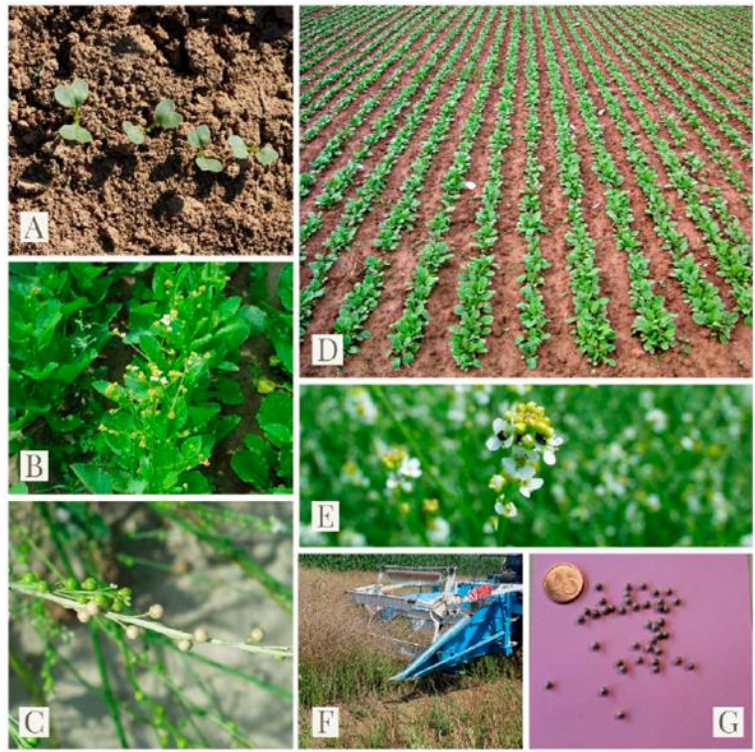

**Fig. 1.** Details of crambe plants: **A**)cotyledons at emergence stage **B**)crambe plants at bolting stage **C**)crambe siliques (hulls) during seed filling stage **D**)crambe at rosette stage **E**)crambe flowers **F**)crambe stand at harvest **G**)crambe seeds [26].

Costa et al. [27] achieved the production of biodiesel using the transesterification method and sodium hydroxide (NaOH) catalyst with the crambe abyssinica oil. They obtained biodiesel from the seeds of the crambe abyssinica plant and reached the conclusion that it meets the EN14214 standards according to the analysis. Rosa et al. [28] produced biodiesel that meets EN14214 standards using the transesterification method and potassium hydroxide (KOH) catalyst. Gülüm et al. [29] produced biodiesel from corn oil using potassium hydroxide (KOH) and sodium hydroxide (NaOH) catalysts. It has been observed that the biodiesel fuel produced with both catalysts meets EN 14214 and ASTM D 6751 standards. Chung 2010 [30], has studied about biodiesel production from Camellia japonica and Vernicia fordii seed oils via alkali catalysts transesterification. It is achieved 97.7% and 96.1% FAME contents with KOH catalyst from C. japonica and V. fordii seed oils respectively. On the other hand, while the C. japonica seed oil denotes high activity with NaOH, but V. fordii seed oil FAME content is insufficient level. Rashid et al. 2008 [31] have obtained biodiesel from sunflower oil with KOH and NaOH catalysts. They have stipulated optimum circumstances for methanolysis of sunflower oil as followed; methanol/sunflower oil molar ratio, 6:1, reaction temperature, 60 °C, reaction time 120 min, stirring rate 600rpm, 1.00% (w/w) NaOH and KOH catalysts. Sunflower oil methyl ester yields for NaOH catalyst and KOH catalyst were attained of 97.1% and 86.7% respectively. Dias et al. 2008 [32] have rummaged about biodiesel yield (wt%) using 0.4 (wt%) NaOH and 0.4 (wt%) KOH catalysts. Higher biodiesel yields using that of KOH and NaOH catalysts have been notched up from soybean oil, sunflower oil and waste frying oil respectively. Vicente et al. [33] have scrutinized different basic catalysts (sodium methoxide, potassium methoxide, sodium hydroxide and potassium hydroxide) for biodiesel production from sunflower oil in their study. Biodiesel yield of 91.67 wt% with using potassium hydroxide while biodiesel yield of 86.71 wt% with sodium hydroxide, have occurred. Singh et al. [34] have assessed potassium methoxide, sodium methoxide, potassium hydroxide and sodium hydroxide catalysts effect on biodiesel yield on 3-level and 4-factor experiments. Potassium methoxide and potassium hydroxide catalysts have established superiority than sodium methoxide and sodium hydroxide catalysts in biodiesel yield:

$KOCH_3 > NaOCH_3 > KOH > NaOH$

Huang et al. [35] have reconnoitered biodiesel deriving from Tra and Basa catfish fat using NaOH and KOH catalysts. Acquired results are as follows: The maximum transesterification yield for Tra fat 89.9% and 91.3% was occurred with NaOH concentration of 0.8% and KOH concentration of 0.8% respectively, whereas maximum yield for Basa fat of 96.3% and 97.1% was come about with NaOH concentration of 0.5% and KOH concentration of 0.8% respectively. Mahlia et al. [36] in their detailed patent research on biodiesel have realized that biodiesel production by transesterification method is simpler and lower costs compared to other methods. Rajkumar et al. [37] aimed to reduce $NO_x$ emission by using biodiesel in diesel engines. They investigated the performance, combustion and emission characteristics of the diesel engine using experimental and modeling methods, using different combinations of biodiesel fuels produced from Karanja and coconut oils. They concluded that $NO_x$ emission decreased with the increase in the ratio of fuel produced with coconut in the biodiesel blend. Karanja has presented a reduction of approximately 18% in $NO_x$ emission with the hydrogenation process of biodiesel fuel. Abed et al. [38] investigated the effect of biodiesel fuels produced from jatropha, palm, algae and waste cooking oil on the values of CO, $CO_2$, $NO_x$, HC and smoke emissions released as a result of the use of the diesel engine. They have tested biodiesel fuels at different concentrations such as B10 and B20 on a single-cylinder diesel engine under different load conditions. In all biodiesel fuels, CO, HC and smoke emissions were found to be lower than standard diesel fuel. It has been concluded that biodiesel fuel produced from waste cooking oil emits more $CO_2$ emission in B10 and B20 concentrations than other biodiesel and standard diesel fuels. Uyumaz et al. [39] produced biodiesel fuel from waste tyre. They investigated the effects of W10 (waste tyre oil fuel blend) biodiesel fuel on the single-cylinder diesel engine, in-cylinder pressure, ignition delay, and combustion time and engine performance. As a result of the tests, it was concluded that using W10 biodiesel fuel instead of standard diesel fuel causes higher in-cylinder pressure and heat transfer. Although W10 biodiesel fuel used on diesel engine showed performance values close to standard fuel, they observed an increase in BSFC (brake specific fuel consumption) by 18.5% at 11.25 Nm engine load. Uyumaz [40] investigated the effect of L10, L20 and L30 fuel mixtures, which contain linseed biodiesel fuel and diesel fuel mixture, on engine performance, specific fuel consumption and emission values. The usage of biodiesel fuel blends has revealed reductions on CO and soot emissions. They concluded that if the engine was loaded with 18.75 Nm and the L30 fuel was used, the CO emission decreased approximately 36.2%, whereas $NO_x$ emission increased by 12.7%. They evaluated it as L10 as the most suitable fuel. Uyumaz et al. [41] carried out detailed performance tests and combustion analysis of OP10 and OP20 mixture of biodiesel produced from poppy oil on a single cylinder diesel engine at 2200 rpm and different engine loads. In the test results, it was observed that biodiesel blends increased the in-cylinder pressure and heat release rate amount. Under full load conditions of the engine, $NO_x$ emission increased by 2.9% and 5.98%, CO emission decreased by 14% to 17.2% with OP10 and OP20 biodiesel fuels compared to diesel fuel, and thermal efficiency was 5.73% and 13.05% found that it decreased.

In the literature, there is limited number of studies on the engine performance, combustion and emission values of biodiesel fuel produced from crambe abyssinica plant involving KOH and NaOH catalysts. It is considered that the current study is aimed to clarify this issue. In this study, the effects of biodiesel fuel produced from crambe abyssinica plant using KOH and NaOH catalysts on engine performance, combustion and emission values were investigated by mixing with standard diesel fuel.

**2 Materials and methods**

**2.1 The engine test setup and specifications**

Test fuels consisting of different biodiesel concentration and standard diesel were examined on a test setup containing a single cylinder diesel engine. Test setup basically consists of diesel engine, dynamometer (Cussons p8160 regenerative D.C. dynamometer), emission device, smoke meter, precision balance, thermocouple, data processing card and computer. In addition to these, a sensitive pressure sensor is used to measure the in-cylinder pressure and an encoder capable of measuring 0.36 °CA precision in determining the engine crank angle. The schematic representation of the test setup where the experiments are carried out is shown in Fig. 2.

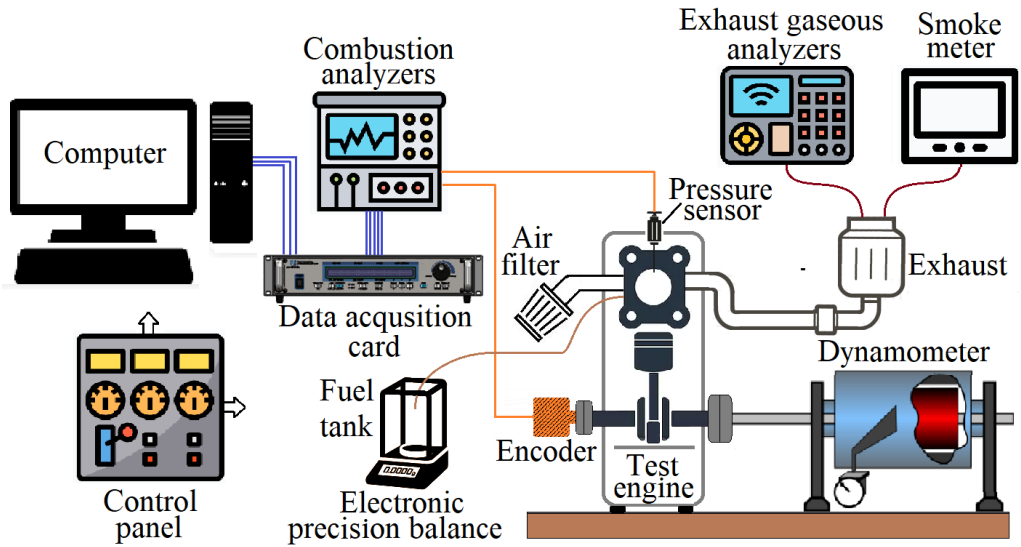

**Fig. 2.** Schematic representation of the test setup

All fuels were tested at 3.75, 7.5, 11.25 and 15 Nm load. The technical specifications of the test engine are shown in Table 1. No structural changes were made to the diesel engine during the test processes.

**Table 1**

Test engine specifications

| Model | Antor/6LD400 |
|---|---|
| Engine type | Direct injection, naturally aspirated |
| Number of cylinder | 1 |
| Bore x Stroke [mm] | 86 x 68 |
| Cylinder volume [cm$^3$] | 395 |
| Compression ratio | 18:01 |
| Maximum power [kW] | 5.4 @ 3000 rpm |
| Maximum torque [Nm] | 19.6 @ 2200 rpm |
| Geometry of combustion chamber | U type |
| Fuel injection system | PF Jerk type fuel pump |
| Injection nozzle | 0.24 [mm] x 4 hole x 160° |
| Nozzle opening pressure [bar] | 180 |
| Injection timing [°CA] | 24 before top dead center (BTDC) |

**2.2 Test Fuels**

By using KOH and NaOH catalysts, biodiesel fuels produced from crambe abyssinica plant were mixed with standard diesel fuel and fuels of different concentrations were obtained. Standard diesel fuel is called D0, the mixture of biodiesel obtained with KOH catalyst and standard diesel fuel is CAK, and the mixture of biodiesel obtained with NaOH catalyst and standard diesel fuel is CAN. Transesterification parameters are as follows 6:1 Methanol to oil molar ratio, KOH catalyst 0.8 g (w/w), NaOH catalyst 0.40 g (w/w), 57°C reaction temperature, and 60 min reaction time. For separation of the methyl esters and glycerol, the mixture was left in separating funnel for 8–10 h. After removing glycerin from the separating funnel, obtained sample was washed with hot distilled water (about 85°C) several times until the washing water became clear. Drying procedure of biodiesel was occurred at 120°C for 20 min to remove any remaining water.

Fatty acid composition of crambe abyysinica oil was shown in Table 2 [42]. The degree of unsaturation of biodiesel was figured out using Equation 1 from the weight ratios of mono and polyunsaturated fatty acids in the fatty acid composition [43-44].

$$DU = (monounsaturated\ C_n: 1, wt\%)$$
$$+2(polyunsaturated\ C_n: 2, wt\%)$$
$$+3(polyunsaturated\ C_n: 3, wt\%) \quad (1)$$

The long chain saturation factor (LCSF), which is considered as a measure of the behavior of biodiesel fuel at low temperatures, was calculated from fatty acids using Equation 2. [45].

$$LCSF = (0.1 \times C16) + (0.5 \times C18) + (1 \times C20) + (1.5 \times C22) + (2 \times C24) \quad (2)$$

The saponification number (SN) refers to the amount in milligrams of potassium hydroxide (KOH) required to saponify 1 gram of oil. The iodine value (IV) is the weight, in grams, of iodine required to saturate the double bonds in 100 g of oil and is an indicator of the degree of unsaturation. Saponification number (SN) was determined using Equation 3, iodine value (IV) was calculated using Equation 4. $A_i$, D and $MW_i$ represent the percentage of each ingredient, number of double bonds and molecular mass of each ingredient respectively. The cetane number (CN) was counted up using Equation 5 [46].

$$SN = SUM\left(\frac{560 \times A_i}{MW_i}\right) \quad (3)$$

$$IV = SUM\left(\frac{254 \times D \times A_i}{MW_i}\right) \quad (4)$$

$$CN = \left(46.3 + \left(\frac{5458}{SN}\right) - (0.225 \times IV)\right) \quad (5)$$

Higher heating value was estimated using Equation 6 [47].

$$HHV = 74.468 - 0.0382\rho \quad (6)$$

Flashpoint was reckoned using Equation 7 [48]. $x_P, x_S, x_O, x_{LI}, x_{LN}$ and $x_E$ symbolize the palmitic, strearic, oleic, linoleic, linolenic and erucic acids by wt%, respectively.

$$FP\ (°C) = 205.226 + 0.083x_P - 1.727x_S - 0.5717x_O - 0.3557x_{LI} - 0.467x_{LN} - 0.2287x_E \quad (7)$$

**Table 2**

Crambe abyysinica faty acid composition

| Faty acid composition of crambe abyysinica | | |
|---|---|---|
| C16:0 | Palmitic | 2.07 |
| C18:0 | Stearic | 0.7 |
| C18:1 | Oleic | 18.86 |
| C18:2 | Linoleic | 9 |
| C18:3 | Linolenic | 6.85 |
| C20:0 | Arachidic | 2.09 |
| C22:0 | Behenic | 0.8 |
| C22:1 | Erucic | 58.51 |
| C24:0 | Linoceric | 1.12 |
| Saturated | | 6.78 |
| Monounsaturated | | 77.37 |
| Polyunsaturated (2) | | 9 |
| Polyunsaturated (3) | | 6.85 |
| Total | | 100 |
| Degree of Unsaturation (DU) | | 115.92 |
| Long Chain Saturated Factor (LSCF) | | 6.09 |
| Saponification Number (SN) | | 171 |
| Iodine Value (IV) | | 98 |

Fuels are named D0 (100% diesel), CAK B25 (biodiesel produced with 25% KOH catalyst + 75% diesel), CAK B50, CAK B75, CAN B25 (biodiesel produced with 25% NaOH catalyst + 75% diesel), CAN B50 and CAN B75. The information of the fuel mixtures used in the test process is given in Table 3. The experiments were carried out under 2200 rpm and 3.75, 7.5, 11.25, and 15 Nm load conditions in a diesel engine. During the experiments, the measurements of the in-cylinder pressure were performed depending on the crank angle. Emission values were directly measured with different fuel mixtures at different engine load conditions. The obtained in-cylinder pressure data were evaluated using mathematical equations in MATLAB / Simulink environment by obtaining parameters such as heat release rate, amount of heat transfer from the cylinder, RI, thermal efficiency, imep, ID, start and end of combustion positions.

**Table 3**

Volumetric fractions of fuels

| Test fuels | Biodiesel ratio (%) | Standard diesel ratio (%) | Catalyst |
|---|---|---|---|
| D0 | 0 | 100 | - |
| CAK B25 | 25 | 75 | KOH |
| CAK B50 | 50 | 50 | KOH |
| CAK B75 | 75 | 25 | KOH |
| CAN B25 | 25 | 75 | NaOH |
| CAN B50 | 50 | 50 | NaOH |
| CAN B75 | 75 | 25 | NaOH |

The properties of CAK, CAN and standard diesel fuel are shown in Table 4. Among the values in Table 4, density, viscosity, water and sulfur ratio values were obtained from the analysis result. 6 different fuels with 25%, 50% and 75% concentration of CAN and CAK biodiesel fuel were created with standard diesel fuel.

**Table 4**

The properties for standard diesel, CAK and CAN biodiesel fuels

| Test fuels | CAK | CAN | Diesel | EN 14214 Lower limit | Upper limit |
|---|---|---|---|---|---|
| Density @15 °C (g/cm$^3$) | 879.70 | 880.03 | 841.75 | 860 | 900 |
| Kinematic viscosity (40 °C mm$^2$/s) | 6.492 | 6.716 | 3.354 | 3.5 | 5.0 |
| Water content (ppm) | 378.49 | 403.58 | – | – | 500 |
| Sulfur (mg/kg) | 1.3 | 1.3 | – | – | 10 |
| Cetane number | 56.17 | 56.17 | 53.1 | 51.0 | – |
| Flash point (°C) | 173.6 | 173.6 | 55 | > 101 | – |
| Higher Heating Value | 40.86 | 40.85 | 45.22 | – | – |

7 different fuel mixtures were tested in the test setup. The engine was brought to operating temperature for each experiment, the necessary controls were made and the conditions were kept stable for all experiments in order to enhance reliability.

**2.3 Measurement of exhaust gas emissions**

Exhaust emissions were measured with EGAS-2M model exhaust gas analyzer produced by Environment SA firm. The technical specifications and measurement sensitivity of gas analyzer are shown in Table 5.

**Table 4**

Technical specifications for EGAS-2M emission measuring device

| Model | Graphite 52M | Topaze 32M | MIR 2M |
|---|---|---|---|
| Gases | THC | NO-NOx | CO-$CO_2$-O |
| Measurement procedure | HFID | HCLD | NDIR-paramagnetic |
| Linearity | < %1 | < %1 | < %1 |
| Measuring range | 0-10/30000 ppm | 0-10/10000 ppm | 0-500/10000ppm (CO) |

|  |  |  | 0-1/20 % ($CO_2$) |
|  |  |  | 0-5/25 % ($O_2$) |
| The lowest measurement | 0.05 ppm | 0.1 ppm | <2% (For full scale) |
| Response time | <1.5 | <2 s | <2 s |

During the experiments, smoke was measured with the AVL DiSmoke 4000 brand / model smoke meter whose technical specifications are given in Table 6.

**Table 6**

Smoke meter specifications

| Device model | AVL DiSmoke 4000 | |
| --- | --- | --- |
| Measurement method | Partial flow | |
|  | Opacity | K value |
| Operating range | 0–100% | Accuracy 0.1% |
| Accuracy [$m^{-1}$] | 0–99,99 | 0,01 |

**2.4 Combustion analysis measurement system and calculation method**

During the experiments, in-cylinder pressure signal covering 50 engine cycles with a sensitivity of 2000 pulse/cycle was received. The received signals are transformed into pressure data with a data processing card and transferred to the computer. In order to increase the precision and accuracy of the test results, the average of 50 cycles was evaluated and the evaluation was made over a single cycle.

The measurement of the in-cylinder pressure on the diesel engine is provided by the pressure sensor whose technical specifications are given in Table 7.

**Table 7**

Pressure sensor specifications

| Model | AVL 8QP500c quartz |
| --- | --- |
| Operating range [bar] | 0-150 |
| Measurement precision [pC/bar] | 11,96 |
| Linearity [%] | ±0.6 |
| Natural frequency [kHz] | >100 |

As a result of the tests, engine in-cylinder pressure values depending on the crankshaft angle were obtained for all fuel types. Parameters such as combustion, heat release rate, combustion stages, thermal efficiency, imep, ID, RI and specific fuel consumption are calculated in MATLAB / Simulink environment with the obtained test data.

One of the most important factors in the examination stages of fuel characterization is the heat release rate behavior during the combustion process. The heat release rate calculation in the combustion process depending on the crank angle is provided by Eq. 8, derived from the first law of thermodynamics. In this equation, leaks are neglected, heat transfer from the cylinder walls ($Q_{heat}$) is included [49]. Here, $Q$ is the heat release rate, $P$ is the pressure inside the cylinder, $V$ is the cylinder volume, $k$ is polytropic index and $\theta$ is the crankshaft angle [50].

$$\frac{dQ}{d\theta} = \frac{k}{k-1} P \frac{dV}{d\theta} + \frac{1}{k-1} V \frac{dP}{d\theta} + \frac{dQ_{heat}}{d\theta} \tag{8}$$

Calculation of the heat transfer from the cylinder walls in the heat transfer equation is given in Eq. 9 and its unit is defined as J/°CA. In the equation, $n$ is the engine speed, $h_c$ is the heat convection coefficient, $T_g$ is the mean gas temperature inside the cylinder and $T_w$ is the combustion chamber wall temperature.

$$\frac{dQ_{heat}}{d\theta} = \frac{1}{6n} h_c A_2 (T_g - T_w) \tag{9}$$

For the calculation of the heat conduction realized in the combustion chamber, the equation (W/m²K) created by Woschni seen in Eq. 10 was taken as reference [51]. The $c_m$ in the equation denotes the average piston speed.

$$h_c = 130 V^{-0,06} P^{0,8} T_g^{-0,4} (c_m + 1,4)^{0,8} \tag{10}$$

The net work performed throughout the cycle is calculated in Eq. 11 and the Indicated mean effective pressure is calculated with Eq. 12 [52]. Equivalent $V_{stroke}$ refers to the cylinder volume.

$$W_{net} = \int_{CA0^o}^{CA720^o} PdV \tag{11}$$

$$\text{Imep} = \frac{W_{net}}{V_{stroke}} \tag{12}$$

The equation where thermal efficiency is calculated by using experimental data is shown in Eq. 13. When using standard diesel and biodiesel fuel mixtures as test fuel, two different fuel amounts ($m_{D0}, m_{Bn}$) and lower calorific value ($Q_{D0}, Q_{LH-Bn}$) are entered into the divisor index due to the difference in the lower heating values of the fuel.

$$\eta_T = \frac{W_{net}}{m_{D0}Q_{LH-D0} + m_{Bn}Q_{LH-Bn}} \tag{13}$$

One of the most important issues to be examined in diesel engines and fuels is RI. Eq. 14 gives the intensity of ringing encountered by the engine during one cycle. $T_{max}$ and $P_{max}$ in 7 in the equation represent the maximum in-cylinder temperature and pressure, and $\gamma$ represent the polytrophic coefficient values.

$$RI = \frac{1}{2\gamma} \frac{\left(\beta\left(\frac{dP}{dt}\right)_{max}\right)^2}{P_{max}} \sqrt{\gamma.R.T_{max}} \tag{14}$$

The model of each subsystem was created by providing the input of the equations between Eq. 8 and Eq. 14 to the MATLAB / Simulink simulation program environment. The visual of the Simulink model created is given in Fig. 3. All analyzes were carried out depending on the crank angle. Graphics were created by exporting the results obtained in MATLAB / Simulink environment to an external file.

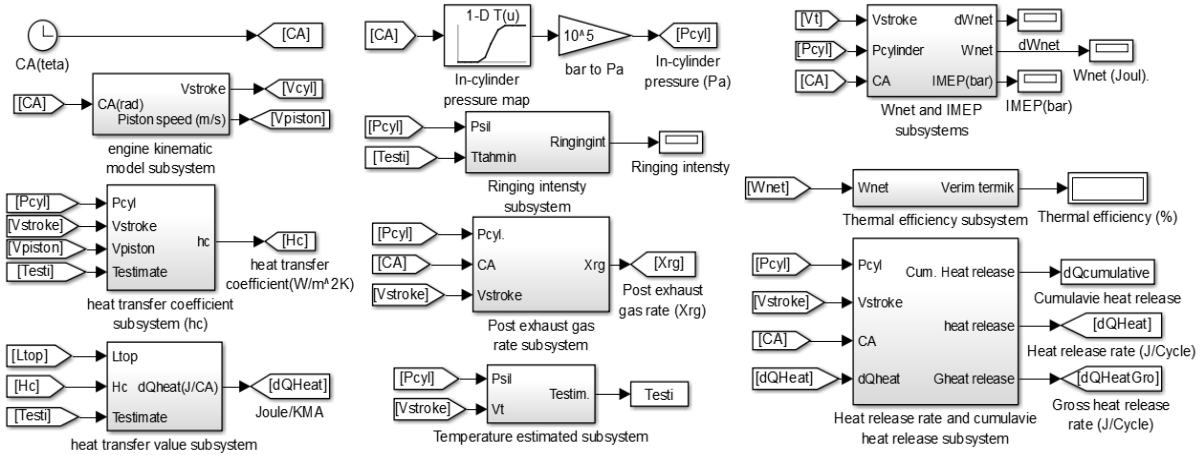

**Fig. 3.** Simulink model visual

**3 Results and Discussion**

The heat release rate obtained by burning the standard diesel fuel under the condition of 3.75 Nm engine torque, in-cylinder pressure, first and second derivative graph and the start of combustion with the start of injection are given in Fig. 4. Fuel injection advance is 24 °CA and injection starts at 336 °CA. As the fuel combustion start, it is determined by accepting the crankshaft angle that the heat release rate changes from negative value to positive value.

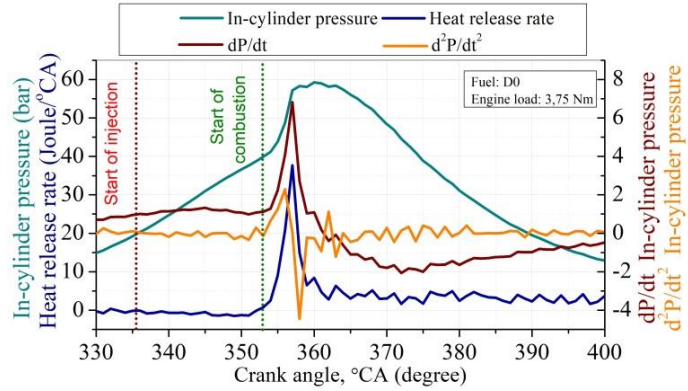

**Fig. 4.** Heat release rate, in-cylinder pressure, 1st and 2nd derivative graphs at 3.75 Nm torque conditions

Combustion stages of standard diesel fuel under 3.75 Nm engine torque condition are shown in Fig. 5. CA10, CA50 and CA90 mean that 10%, 50% and 90% of the charge mixture that is burned versus crank angle. Since the combustion after CA90 point cannot be well detected due to heat transfer to the cylinder wall and after burning process in combustion studies, and it is accepted as the end of combustion.

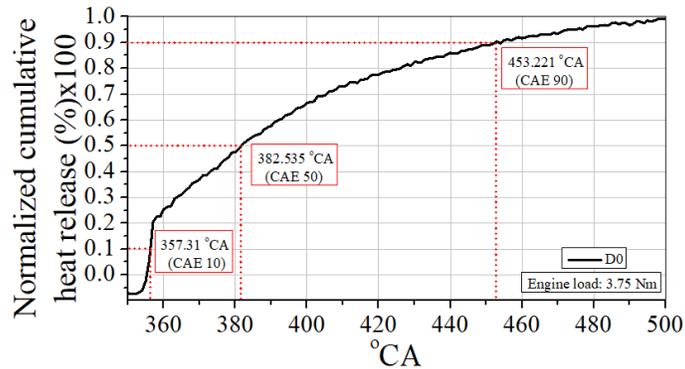

**Fig. 5.** Standard diesel fuel combustion stages at 3.5 Nm torque conditions

The graphics containing the in-cylinder pressure and heat release rate values obtained by burning D0, CAK and CAN fuels at different engine loads are given in Fig. 6. The highest pressure and heat release rate value was obtained with standard diesel fuel when the engine load was 3.75, 7.5 Nm and with CAK B25 mixed fuel at 11.25 and 15 Nm engine load. Under all engine load conditions, it was observed that the maximum cylinder pressure values gradually decreased as the biodiesel concentration rate increased. It was determined that with the same concentration of CAK fuel series, lower maximum pressure values were achieved compared to the CAN fuel series. It was seen two stages of combustion called pre-combustion and diffusion combustion as seen in heat release rate variation. It can be found from Figure 6 that lower heating value caused to obtain lower in-cylinder pressure and heat release with the addition of biodiesel. It can be also mentioned that biodiesel showed better performance with the increase of engine load in view of in-cylinder pressure and heat release rate. Biodiesel cannot be well atomized and vaporized compared that D0 due to higher density and viscosity. It can be explained that longer combustion process showed positive effect with the combustion of biodiesel fuel blends in view of complete combustion.

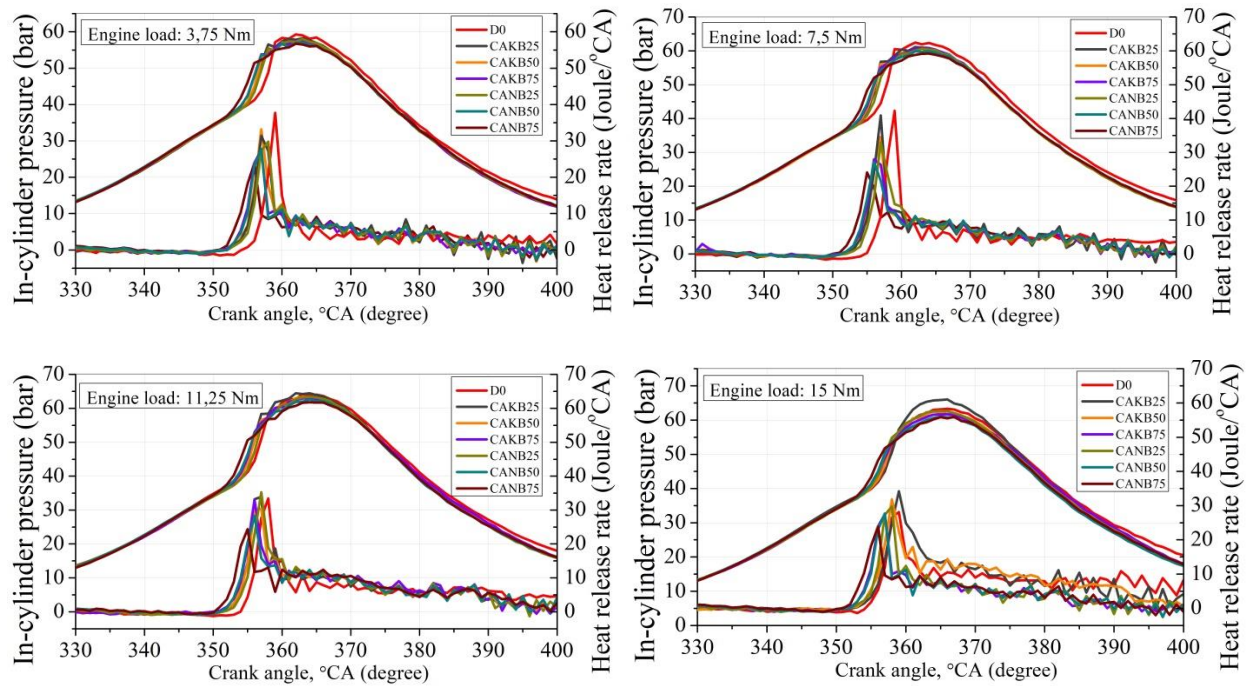

**Fig. 6.** Cylinder pressure and heat release rate graphs at 2200 rpm speed and different load engine conditions

There is significant relationship between thermal efficiency and CA50. If CA50 is obtained nearly after top dead center (ATDC), higher thermal efficiency can be observed. Thermal efficiency and CA50 values determined in different fuel mixtures and engine loads are shown in Fig. 7. It was concluded that the highest thermal efficiency values were achieved with CAK B25 fuel blend at medium and high loads. While the highest thermal efficiency value was obtained in fuel types with a biodiesel concentration of 25%, a gradual decrease was observed in fuel types with a 50% and 75% concentration. The highest thermal efficiency was computed as 21.32% with CAK B25 at 11.25 Nm engine load. It has been determined that the use of standard diesel fuel has higher thermal efficiency than CAK B75 and CAN B75 fuel mixtures under all engine load conditions due to higher heating value. In fact, the excess amount of oxygen in the biodiesel fuel is expected to increase the thermal efficiency. However, it is thought that the thermal efficiency of biodiesel fuel is lower than standard diesel fuel due to the lower heating value and the amount of water in it. It was seen that there is good agreement between thermal efficiency and CA50.

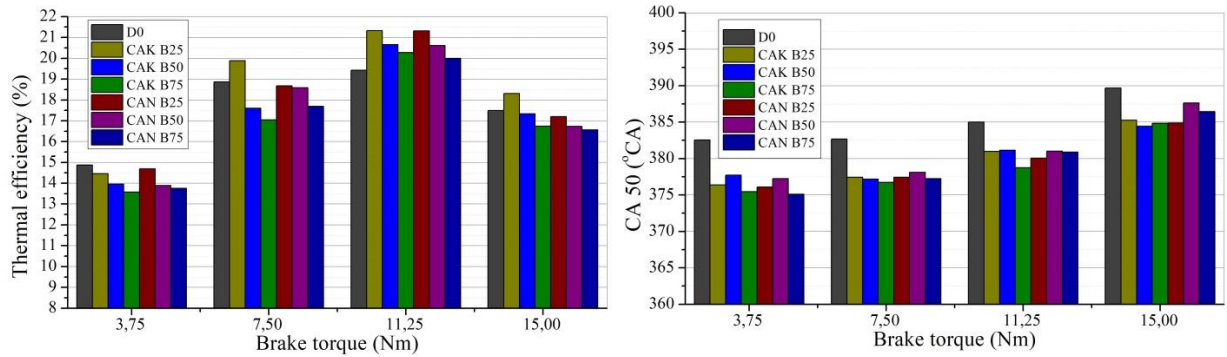

**Fig. 7.** Thermal efficiency and CA50 values at 2200 rpm speed and different load motor conditions

The imep values obtained as a result of the experiment with different fuel mixtures and engine loads are shown in Fig. 8. Imep can be defined as engine performance indication. It was observed that the highest imep value was obtained in CAK B25 as 4.49 bar at 15 Nm engine load. It was found that the same concentration CAK biodiesel fuel mixture had a higher imep value at 3.75, 11.25 and 15Nm engine loads than CAN fuel mixtures. More charge mixture is taken to the cylinder resulting in higher heat release. Another reason for this result is that the water concentration in the CAN fuel mixture is higher than the water concentration in the CAK fuel mixture. The excess of water concentration causes heat absorption and a decrease in pressure during the evaporation process. So, the in-cylinder pressure exerted on the piston in a cycle increases. Hence, imep increases. The lower calorific value of biodiesel fuel is lower than that of standard diesel fuel. Increasing the biodiesel fuel ratio in the fuel mixture causes a decrease in the lower calorific value of the fuel mixture. However, the heat generated during the combustion process decreases. The decrease in the heat generated during the combustion process also causes a decrease in the imep value. On the other hand, the combustion process is improved due to the presence of oxygen in the biodiesel fuel, which causes an increase in imep.

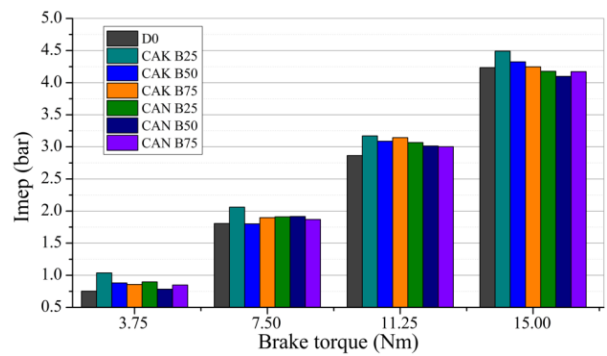

**Fig. 8.** Comparison of imep values at 2200 rpm speed and different load engine conditions

The variations of combustion duration values versus engine load are shown in Fig. 9. Combustion duration was determined between the start of combustion and CA90. It has been observed that the shorter combustion was achieved with standard diesel for all engine load conditions. It was determined that combustion duration decreased with CAK fuel blends at the same concentration according to CAN fuel blends. The longest combustion duration was computed with CAN B75. When standard diesel fuel was used at 3.75 Nm, it was determined that the lowest combustion duration was 96 °CA and the CAN B75 fuel mixture presented the highest combustion duration at 15 Nm. The increase of engine load caused to take more charge mixture resulting in longer time to complete combustion. Higher viscosity and density of biodiesel fuel blends make difficult to atomize and vaporize fuel molecules. Thus, more time is needed to complete combustion.

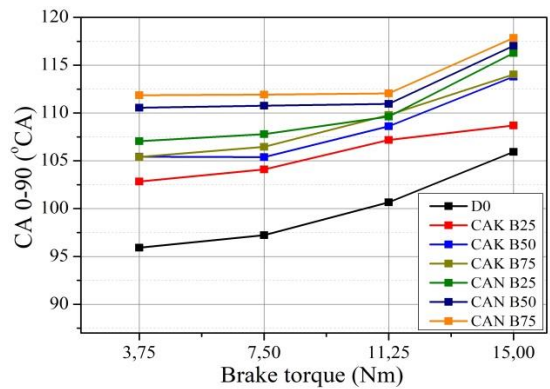

**Fig. 9.** Comparison of CA 0-90 combustion duration at 2200 rpm and different load engine conditions

ID is directly dependent on fuel property that is cetane number. Higher cetane number causes the shorter ID. The time elapsed between the fuel start of injection and the start of combustion is called the ID [53]. ID values obtained by burning D0, CAK and CAN fuels at different engine loads are given in Fig. 10. It has been observed that the ID of the standard diesel fuel is the highest under 3.75, 7.5 and 11.25 Nm torque conditions of the engine, and close to the maximum under 15 Nm load condition. Higher ID is among the expectations that it will cause knock due to the increase of accumulated fuel during combustion. At 11.25 Nm engine load, the lowest ID values were determined. It is observed that the use of CAN fuel mixture increases the ignition delay compared to the use of CAK fuel mixture of the same concentration at higher loads. One of the reasons for this is that the excess water concentration in the CAN fuel mixture increases the resistance to ignition. Surprisingly, biodiesel fuel blends present lower ID compared that diesel fuel in spite of higher density and viscosity of biodiesel. The reason for the low ID value in the use of biodiesel fuel is due to the higher cetane number value of the biodiesel fuel.

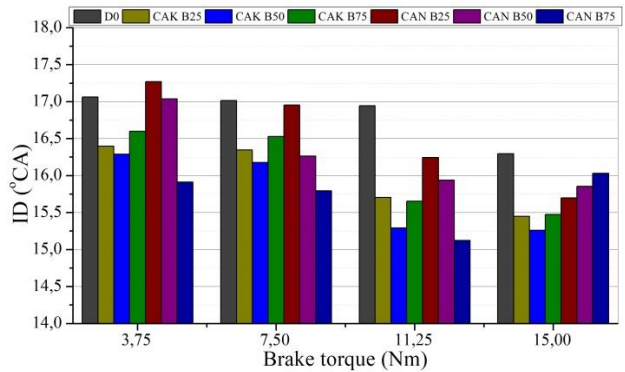

**Fig. 10.** Comparison of ID values at 2200 rpm and different load engine conditions

RI is a combustion parameter depending on engine speed, maximum pressure rise rate. RI values obtained with test fuels at different engine loads are shown in Fig. 11. It is concluded that the use of standard diesel fuel on the engine is more prone to knock than other blended fuels. As the biodiesel ratio increases, the ringing intensity value decreases. This is due to the high cetane number of biodiesel fuel. With the higher cetane number, the ignition of the fuel injected into the cylinder becomes easier and its tendency to knock decreases. The higher ID values of standard diesel fuel compared to other fuels strengthen the accuracy of this expected result. With the increase of biodiesel concentrations, an decrease in the RI of CAK and CAN mixed fuels was observed. It has been observed that CAK fuel blends at the same concentration have lower RI than CAN fuel blends. The highest RI was calculated with DO as 0.03171 MW/m$^2$ at 7.50 Nm engine load. It can be mentioned that RI decreased with the increase of engine load. Combustion of higher charge mixture allowed obtaining higher in-cylinder temperature and pressure. So, combustion conditions are improved across the combustion chamber.

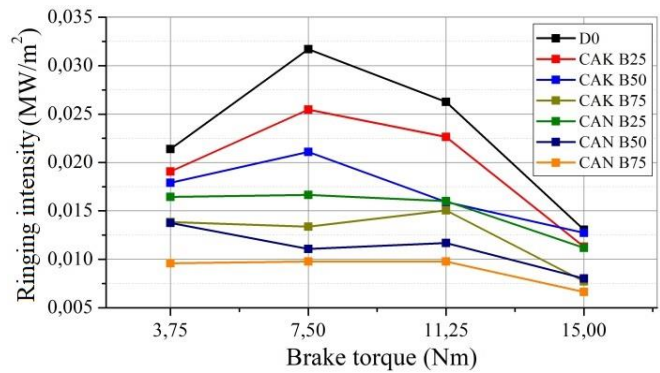

**Fig. 11.** Comparison of ringing intensity values at 2200 rpm and different load engine conditions

Brake specific fuel consumption values obtained as a result of measurements performed at different fuel mixtures and engine loads are shown in Fig. 12. The result is that the lowest BSFC values are provided with standard diesel fuel under all engine load conditions. As the biodiesel concentration of the standard diesel fuel was increased, the BSFC value increased due to lower heating value and higher density. The same concentrated CAN mixed fuel series has been found to have a lower BSFC value at all engine loads than the CAK blended fuel series. The highest BSFC was computed with CAK B75 as 616.691 g/kWh at 3.75 Nm engine load. The lowest BSFC values for all fuel types were achieved at a torque value of 11.75 Nm at the engine load. BSFC first decreased and then started to increase with the increase engine load for all test fuels. At medium engine loads heat losses and gas leakages decrease resulting in lower BSFC.

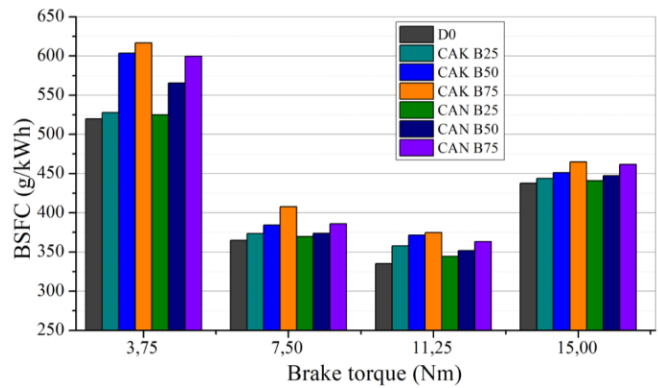

**Fig. 12.** Effect of different biodiesel blended fuels on specific fuel consumption under 2200 rpm speed and different engine load conditions

Incomplete combustion is seen owing to lower oxygen concentration and in-cylinder temperature allowing CO formation that is combustion product. The variations of CO versus engine load is shown in Fig. 13-a. Lower CO was measured at low engine loads because less charge mixture is burned. It can be emphasized that sufficient oxygen molecules exist due to lower fuel concentration in the combustion chamber. Hence, CO formation is reduced. In contrast with this phenomenon, CO increased at higher engine loads because of higher fuel fractions in the charge mixture. This situation caused to deteriorate chemical reactions between fuel and oxygen molecules. So, CO formation is seen again. Biodiesel fuel blends presented lower CO formation due to higher oxygen concentration. Fig. 13-b represents the $NO_x$ variations. $NO_x$ increased with the increase engine load owing to combustion of higher charge mixture. High combustion temperatures are observed at high loads. Oxygen and nitrogen molecules are reacted each other at higher in-cylinder temperature resulting in $NO_x$ formation. The highest $NO_x$ was measured with CAKB75 as

341.656 ppm at 15 Nm engine load. It was also seen that $NO_x$ increased with the increase of biodiesel fraction in fuel blends. The lowest $NO_x$ was measured with diesel for all loads. Minimum $NO_x$ was determined with D0 as 22.489 ppm at 3.75 Nm. In the same conditions, it has been observed that lower NOx emission values are obtained in the case of using CAK fuel instead of using CAN fuel mixture. One reason for this is that the water concentration of the CAN fuel mixture is higher than that of the CAK fuel mixture. The increase in the water concentration in the fuel causes the formation of water vapor in the combustion process and, together with it, a decrease in the combustion end temperature. NOx emission value decreased slightly with the decrease of the combustion end temperature. As can be seen in Figure 13-c, an increase in smoke emissions was observed with increasing engine load. Smoke emissions are generally resulted from rich charge mixture at higher engine loads. Incomplete combustion occurs as more fuel is injected at higher altitudes, resulting in higher smoke levels. The highest smoke was determined as 2.29 $m^{-1}$ and 2.59 $m^{-1}$ for D0 at 11.25 and 15 Nm engine loads respectively. It was observed that there was a decrease in smoke emission as the biodiesel ratio increased. Fuels with biodiesel have a higher oxygen content and less carbon compared to diesel fuel, which explains the reduction in smoke. Similar change was seen with HC emission as shown in Fig. 13-d. HC is emitted due to incomplete combustion because fuel and oxygen molecules could not well react near the cylinder wall and piston cavity in the combustion chamber. The highest HC was determined as 2400.27 ppm with D0 at 15 Nm engine load. It was seen that HC reduced with the usage of biodiesel fuel blends. Oxygen concentration caused to improve oxidation reactions according to D0 with biodiesel. It has been observed that the use of CAN fuel mixture results in higher HC emissions than the use of CAK fuel mixture of the same concentration under 3.75, 7.5 and 11.25 Nm engine load conditions. The main reason for this is that the CAN fuel mixture adversely affects combustion due to the excess water content in the mixture.

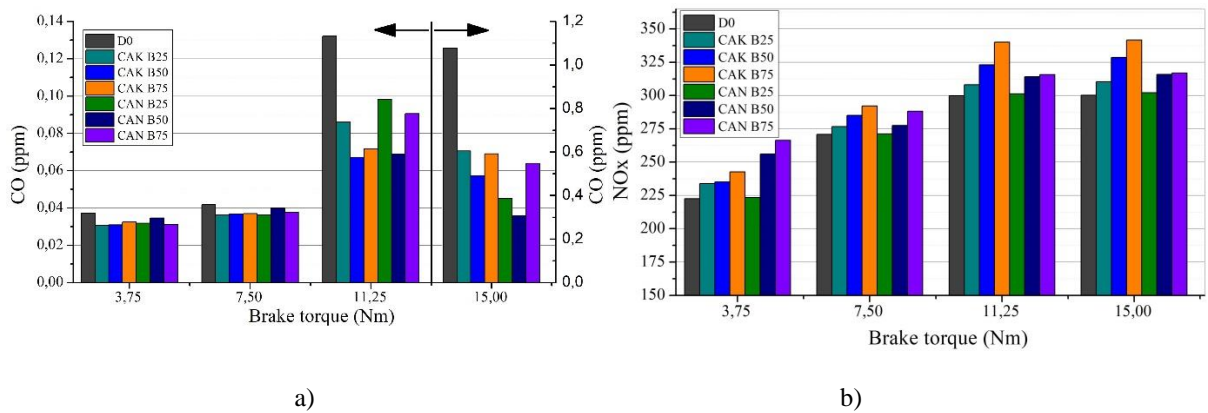

a) b)

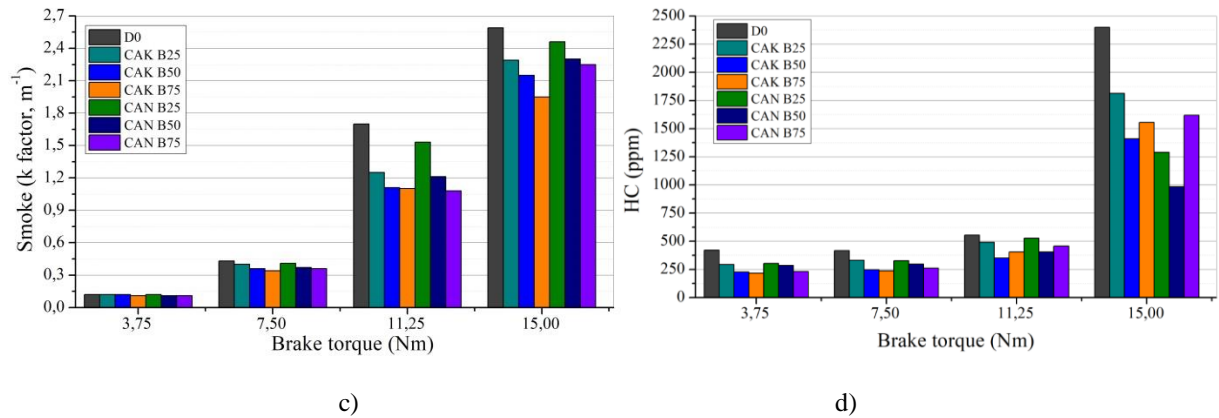

c)                                                                 d)

**Fig. 13.** Effect of different biodiesel blended fuels on emissions under 2200 rpm speed and different engine load conditions

## 4 Conclusion

In this study, the effects of biodiesel fuels produced by using KOH and NaOH catalysts from crambe abyssinica plant, which is in the non-edible plant group, on combustion, engine performance and emission were experimentally investigated. The tests of standard diesel fuel (D0) and 6 different fuel mixtures (CAK B25-50-75 and CAN B25-50-75) were carried out at 2200 rpm engine speed and 3.75, 7.5, 11.25 and 15 Nm engine load conditions. The data obtained as a result of the experiments were analyzed in MATLAB / Simulink environment and parameters such as combustion heat release rate, combustion stages, and thermal efficiency, indicated mean effective pressure, ignition delay, ringing intensity and specific fuel consumption were calculated and evaluated. The study result is summarized as follows:

- It has been proved that the CAN and CAK blended fuel series with 25%, 50% and 75% biodiesel concentration can be used without any structural changes on the diesel engine.

- The highest thermal efficiency values were obtained by using CAK B25 mixed fuel under all engine load conditions. In CAK and CAN mixed fuels, the highest thermal efficiency value was obtained when the concentration was 25%, while a gradual decrease was observed in the fuel types with a concentration of 50% and 75%. It was determined that the use of D0 fuel at all engine load conditions has higher thermal efficiency than CAK B75 and CAN B75 fuel mixtures.

- It was observed that the highest imep value was obtained in CAK B25 fuel at all engine loads. It was found that the same concentration CAK biodiesel fuel mixture had a higher imep value at 3.75, 11.25 and 15Nm engine loads than CAN fuel mixture.

- It was found that the lowest burning time was provided with D0 fuel when the engine load was 3.75 Nm with 96 ºCA, and the highest combustion time was provided with CAN B25 fuel when the engine load was 15 Nm with 116 ºCA.
- It was concluded that the ignition delay of D0 fuel was the highest in the engine's 3.75, 7.5 and 11.25 Nm torque conditions, and close to the maximum under 15 Nm load condition. It is concluded that the use of standard diesel fuel on the engine is more prone to knock than other blended fuels. The high ignition delay values of standard diesel fuel compared to other fuels and being compatible with the literature strengthens the accuracy of this result.
- It has been determined that the BSFC value of D0 diesel fuel is lower than CAN and CAK fuel mixtures under all engine load conditions. The same concentrated CAN fuel series has been found to have a lower BSFC value at all engine loads than the CAK fuel series. The lowest BSFC values for all fuel types were achieved at a torque value of 11.75 Nm at the engine load.
- It is concluded that using CAK and CAN biodiesel fuel mixture on diesel engine instead of standard diesel fuel reduces CO, HC and smoke emission values. In all engine load conditions of CAK and CAN fuel mixtures, as their concentration increases, the $NO_x$ value increases and the smoke values decrease. In the results obtained with the same concentrated CAK fuel series at all engine loads, it was determined that $NO_x$ emissions were lower than the results obtained compared to the CAN fuel series.

**Acknowledgement**

As researchers, we thank the Gazi University Projects of Scientific Investigation (BAP) to supporting this study which was partly in frame of the project code of 07/2011-54.

**Nomenclature**

| | | | |
|---|---|---|---|
| BSFC | Brake specific fuel consumption | $k$ | Polytrophic index |
| CA | Crank angle | KOH | Potassium hydroxide |
| CA10 | Crank angle location of 10% accumulated HRR (°CA) | $n$ | Engine speed |
| CA50 | Crank angle location of 50% accumulated HRR (°CA) | NaOH | Sodium hydroxide |
| CA90 | Crank angle location of 90% accumulated HRR (°CA) | $P$ | In-cylinder pressure |
| CAK | Crambe abyssinica KOH catalyst | $Q$ | Heat release rate |
| CAN | Crambe abyssinica NaOH catalyst | RI | Ringing intensity |
| CI | Compression ignition | SOC | Start of combustion |
| CO | Carbon monoxide | SOI | Start of injection |
| $CO_2$ | Carbon dioxide | $\theta$ | Crank angle |
| HC | Hydrocarbon | $T_g$ | In-cylinder temperature |
| $h_c$ | Heat convection coefficient | $T_w$ | Cylinder wall temperature |
| imep | Indicated mean effective pressure | $V$ | Cylinder volume |
| ID | Ignition delay | | |